# A dual chirped-pulse electro-optical frequency comb method for simultaneous molecular spectroscopy and dynamics studies: Formic acid in the THz region

JASPER R. STROUD[1,*] AND DAVID F. PLUSQUELLIC[1,*]

[1]Applied Physics Division, Physical Measurement Laboratory, National Institute of Standards and Technology, Boulder, CO 80305
*Corresponding authors: jasper.stroud@nist.gov and david.plusquellic@nist.gov

**An electro-optic dual comb system based on chirped-pulse waveforms is used to simultaneously acquire temporally magnified rapid passage signals and normal spectral line shapes from the back-transformation to the time domain. Multi-heterodyne THz wave generation and detection is performed with the difference frequency mixing of two free-running lasers. The method is used to obtain THz spectra of formic acid in the 10 cm$^{-1}$ to 20 cm$^{-1}$ (300 GHz – 600 GHz) region over a range of pressures. The method is widely applicable across other spectral regions for investigations of the transient dynamics and spectroscopy of molecular systems.**

Dual-comb spectroscopy has revolutionized optical sensing by providing a straightforward and precise method for the down conversion and compression of the optical probe comb into the radio frequency (RF) domain to enhance detection sensitivity [1,2]. Electro-optic dual comb methods, while often limited in spectral coverage, are highly flexible across a wide range of optical resolutions and measurement time scales. When both combs are generated from the same laser source, complicated phase locking schemes are often not necessary to achieve transform limited readout performance [3,4].

As opposed to electro-optic frequency comb generation that consists of series of transform limited pulses with flat spectral phase, here we use a series of linear chirped pulses that have a quadratic phase response to form optical frequency combs [5,6]. Linear chirps facilitate the production of flat comb spectra and provide continuous power at the detector that helps alleviate dynamic range issues. In contrast to traditional dual comb setups, the probe and local oscillator (LO) sources have the same repetition rate, or chirp duration, but differ in the chirp range. The relationship between the LO and probe combs can be described as,

$$\Delta f_{LO} = \Delta f \pm \delta f_{BW} \qquad (1)$$

where $\Delta f_{LO}$ is the chirp range of the LO comb, $\Delta f$ is the chirp range of the probe comb, and $\delta f_{BW}$ is the small difference in the chirp range that defines the RF comb bandwidth. This mapping is done when the LO chirp beats with the probe chirp, generating sum and difference frequency components. The detected RF chirp can be described as a cosine waveform that includes a quadratic phase term,

$$RF(t) = I_{RF}(t)\cos\left(\pm 2\pi\left(\delta f_0 t + \frac{\delta f_{BW} t^2}{2\tau_{CP}}\right)\right) \qquad (2)$$

where $\delta f_0$ is some offset frequency between the two combs, $\tau_{CP}$ is the duration of the chirp, and $I_{RF}(t)$ is half the product of the probe and LO field strengths. The quadratic term maps the information from the probe onto an RF bandwidth defined by the difference in chirp rates, $\delta f_{BW}$.

When the chirped waveforms are repeated for a fixed set of LO phase slips [5,6], interleaved RF combs are formed in the frequency domain that are unique to each of the electro-optic modulator's (EOMs) (±) sidebands and orders (harmonics). When the applied chirped field excites a resonance, the resulting transient oscillations no longer have the quadratic phase term of the probe. In contrast to Eq. (2), the transient mixing with the LO chirp can be described as,

$$RF_{trans}(t) = I_{trans}(t)\cos\left(2\pi\left(\delta f_{trans} t + \frac{\Delta f_{LO} t^2}{2\tau_{CP}}\right)\right) \qquad (3)$$

where, $\delta f_{trans}$ is the frequency of the resonance response, and $I_{trans}(t)$ is the product of this transient response and LO field strengths. Here, the quadratic term is defined by the LO chirped pulse and results in the temporally magnified transient oscillations observed in the RF comb spectrum. See the supplemental S1 and S2 document for derivations of Eq. 2 and Eq. 3, respectively.

This scheme has been shown by us to magnify the temporal dynamics of carbon dioxide in the near infrared [5] and of water vapor in the THz region [6]. In this manuscript, we demonstrate this differential chirped-pulse down conversion using a dual-chirped-pulse difference-frequency electro-optical frequency comb system (DCP-DF-EOFC) to measure the THz spectrum of formic acid. The magnified and steady state spectral responses are recovered from this dual comb setup in multiple configurations. We also show how the sign of the quadratic phase shift can be flipped through simple modification of the LO chirped-pulse parameters to effectively reverse the direction of the temporal response in the magnified spectrum.

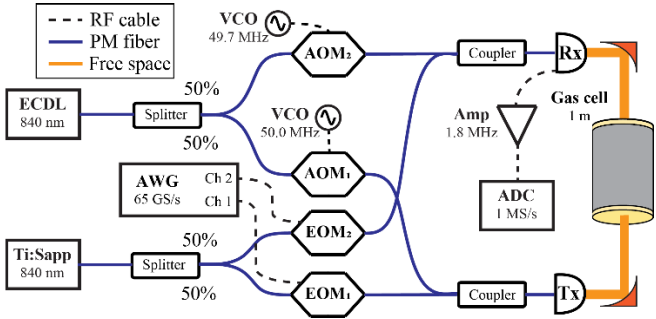

**Fig. 1**. Schematic of the DCP-DF-EOFC spectrometer.

The system diagram is shown in Fig. 1 and described in detail elsewhere [6]. Briefly, two continuous wave lasers, an external cavity diode laser (ECDL) and a Ti:Sapphire (Ti:Sapp) ring laser, are used to generate THz radiation. The laser outputs are coupled into fiber, equally split, and sent to optical modulators (see Fig. 1). The two acousto-optic modulators (AOM) are centered ≈ 50 MHz and differ in driving frequency (VCO) by $\Delta f_{AOM}$ = 300 kHz. The two EOMs are driven by a dual channel arbitrary waveform generator (AWG) programmed with chirped pulse waveforms. The combined fiber outputs with 20-30 mW each illuminate a photomixer transmitter (Tx) and receiver (Rx) [7]. The Tx photomixer (11 V bias) output is free-spaced coupled through a 1 m long gas cell using a pair of gold-coated off-axis parabolic mirrors. The optical pathlengths of each leg are approximately matched at the detector to reduce phase drift. The THz light on Rx is mixed with the optical beat signals to generate an RF electrical response which is amplified and digitized (ADC, 1 megasample/s, MS/s).

The chirped pulse waveform driving the Tx leg spans from $f_{Tx,start}$ = 0.5 GHz to $f_{Tx,stop}$ = 10.5 GHz in a 10 ms duration. The Rx leg is a chirped pulse with the same duration but spans a slightly different bandwidth where $f_{Rx,start}$ = $f_{Tx,start}$ ± $f_{RF,start}$ and $f_{Rx,stop}$ = $f_{Tx,stop}$ ± $f_{RF,stop}$, and $f_{RF,start}$ = 5 kHz and $f_{RF,stop}$ = 164 kHz. The Tx chirped pulse is repeated ten times while each Rx waveform is phase shifted to separate the orders (harmonics) generated in the EOMs [5,6]. The 100 ms long AWG waveforms (32 GS/s) are repeated ten times to form a continuous 1 s record. The resulting 1 s RF interferograms are Fourier transformed to give the RF spectrum with a comb spacing of 100 Hz that spans 159 kHz on each side of the 300 kHz AOM beat note (with a 5 kHz center offset). The comb teeth are sampled [6] and normalized to recover the magnitude and phase of the spectra in the frequency domain (Fourier transform of the recorded interferogram). Prior to normalization to the background signal, the comb teeth are inverse Fourier transformed to obtain the steady-state spectrum here defined as the time domain spectrum. Through the unique Fourier transform properties of linear chirps, the frequency domain and time domain spectra are obtained simultaneously. Normalization removes the quadratic phase term that defines a linear chirp and allows for transformation between the two domains (see supplemental S3 for details).

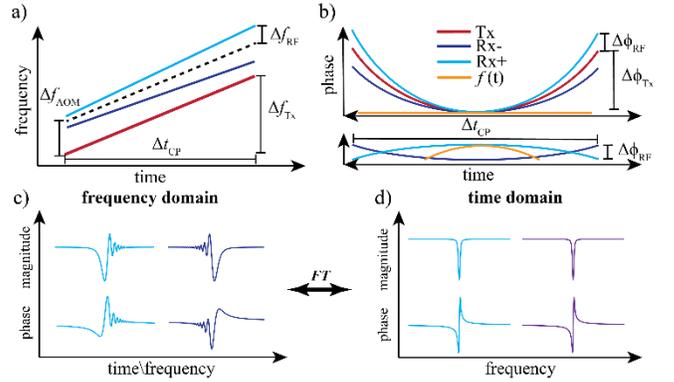

**Fig. 2**. (a) The spectrogram of chirped waveforms illustrating the relationship between the two different Rx waveforms, in purple and blue, and the Tx waveform, in red. (b) The quadratic terms of the chirps are shown in (a), and the arguments in Eqs. 2 and 3 are below. (c) The rapid passage frequency domain spectra of a single resonance, and (d) the normal line shape time domain spectra for both Rx chirps.

To better illustrate the expected forms of the temporally magnified signals, two different LO chirped pulse waveforms were applied to the Rx leg to down-convert the spectrum of the Tx probe comb. Figure 2 shows how for a single resonance, the different Rx chirps interact to produce mirror image frequency domain spectra while having identical time domain spectra. The spectrogram in Fig. 2a shows the Tx chirp in red (and an offset copy of it, dashed black line) and the two Rx chirps, one with a slightly larger chirp ($\Delta f_{LO} > \Delta f$) in blue relative to the Tx chirp, and one with a slightly smaller chirp ($\Delta f_{LO} < \Delta f$), in purple. Figure 2b shows an overlay of only the quadratic terms of the chirps shown in Fig. 2a, the probe in red, and the two Rx chirps in purple and light blue (see supplement S1). The difference in these phases (lower subpanel of Fig. 2b), which are the arguments in Eqs. 2 and 3, describes the mapping of the probe frequency comb to the RF domain. The temporal response of the sample at $\delta f_{trans}$ in Eq. 3, in yellow, acquires the quadratic phase response from mixing with the LO chirp. While the spectral content in the probe's (Tx) chirped pulse is mapped with the differential term given in Eq. 2, the transient response is magnified by the quadratic term of the LO chirped pulse given in Eq. 3. The normalized frequency domain response from a sample resonance, shown in Fig. 2c, is found from,

$$FD(t) = C_{Sig}(n)/C_{Bkg}(n) \qquad (4)$$

where $C_{Sig}(n)$ and $C_{Bkg}(n)$ are the complex values of the signal and background combs, respectively. The direction of the decaying oscillations (ripples) depends on the sign of the RF quadratic phase response (Fig. 2b). However, when the inverse Fourier transform (iFT) is applied prior to normalization, the steady state (i.e., normal line shape) spectra in the time domain are recovered as follows,

$$TD(f) = iFT\{C_{Sig}(n)\}/ iFT\{C_{Bkg}(n)\} \qquad (5)$$

These two spectra are nearly identical for both Rx combs as illustrated in Fig. 2d. The temporally magnified frequency comb spectra contain distinct post-resonance oscillations that are interpreted as the rapid passage responses of the molecular resonance [5,6,8] (see supplement S4). Therefore, the frequency domain spectra have a temporal response in contrast to the frequency response of the time domain spectra. These concepts are demonstrated here through investigations of the rapid passage dynamics and spectroscopy of formic acid in the THz region.

The DCP-DF-EOFC system is used to perform gas phase spectroscopy of formic acid. Three spectral regions are investigated in this work that are centered at 12.1 cm$^{-1}$, 15.7 cm$^{-1}$, and 19.3 cm$^{-1}$ (362 GHz, 472 GHz, and 581 GHz, respectively). Over these spectral regions, the down-converted THz signal strength and corresponding signal-to-noise ratio varies by ≈ four-fold. For each region, two sets of data were collected at 133 Pa, 267 Pa, 400 Pa, and 667 Pa, using two different chirped pulses for the Rx comb. Each data set produces four spectra that include the magnitude and phase responses in both the frequency and time domains. The frequency domain spectra show rapid passage line shapes that converge to normal Voigt profiles with increasing pressure, while the time domain spectra always show normal Voigt profiles having regular pressure-broadened widths.

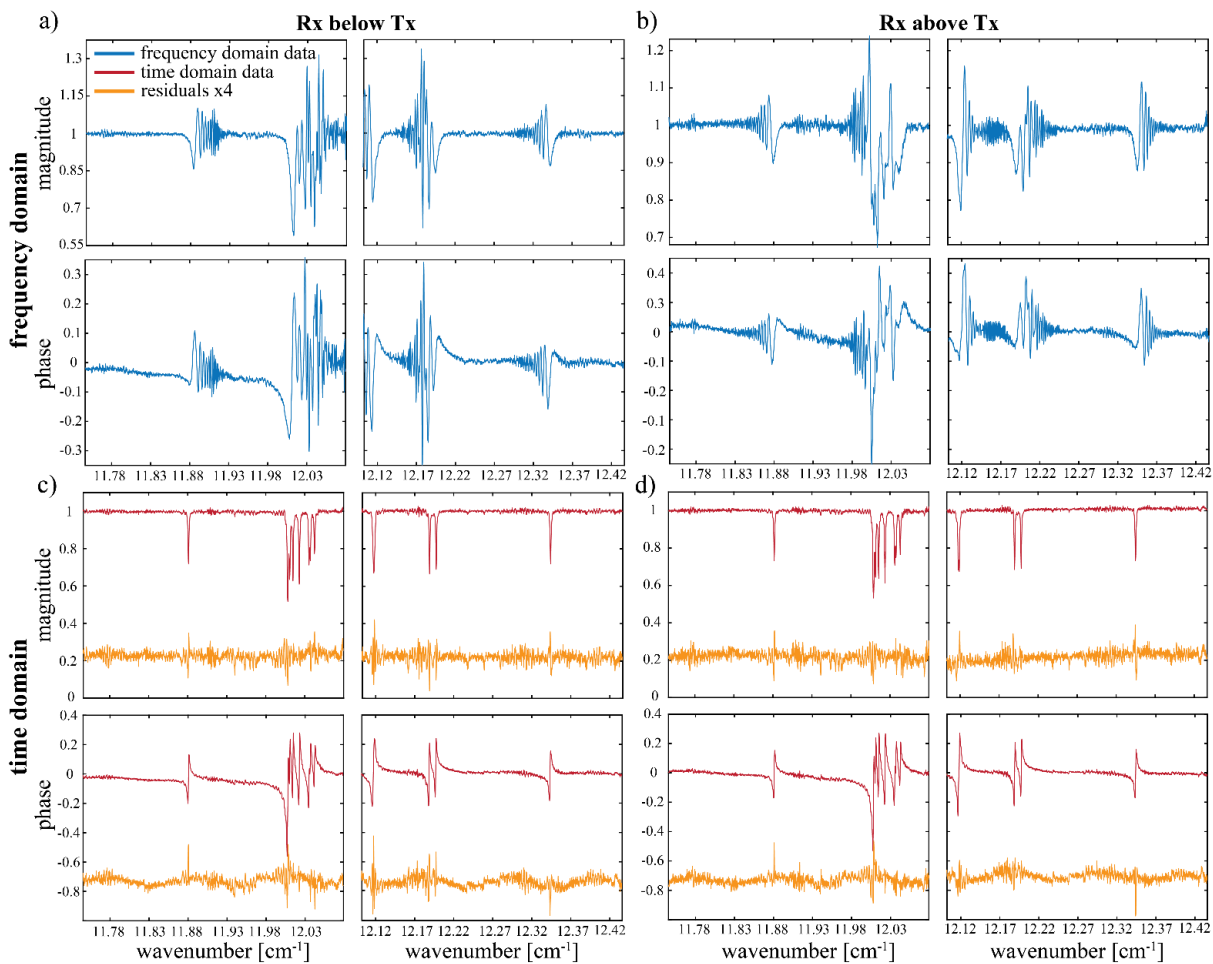

**Fig. 3**. The magnitude and phase spectra of formic acid (≈ 95% purity) around 12.1 cm$^{-1}$ in a 1 m long gas cell at 133 Pa pressure. (a) The quadratic phase response from the LO chirp causes the ripples to move towards the center frequency (between sub-panels). (b) The quadratic phase response causes the ripples to move away from the laser center frequency. (c) The time domain (steady state) spectra in red correspond to the temporally magnified spectra in (a). (d) The steady state spectra in red correspond to the temporally magnified spectra in (b). The residuals from the fits to the HITRAN database are shown in orange (offset for clarity) in (c) and (d).

For the first data set, the ECDL was tuned to ≈ 11866 cm$^{-1}$, while the Ti:Sapp was fixed at ≈ 11878 cm$^{-1}$ to generate a difference frequency THz comb centered at 12.08 cm$^{-1}$. Following the empty cell acquisition, the 1 m long evacuable gas cell was filled with up to 667 Pa of ≈ 95% formic acid (residual 5 % water vapor). Figure 3 shows the resulting frequency and time domain spectra obtained at room temperature and 133 Pa pressure for the two different Rx chirped pulses (see Fig. 2a). The frequency domain spectra shown in Fig. 3a and 3b contain oscillations from each resonance from rapid passage effects. Figure 3a shows the spectra for $\Delta f_{LO} < \Delta f$, resulting in ripples that are dampened toward the comb's center frequency. In contrast, for $\Delta f_{LO} > \Delta f$ in Fig 3b, the ripples to extend away from the comb's center frequency. Since this region contains multiple lines of formic acid, the responses significantly overlap when temporally magnified, and therefore, exhibit distinct spectra depending on the direction of the rapid passage signals.

However, after applying the inverse Fourier transformation to the time domain as described above, the corresponding steady state spectra shown in Figs. 3c and 3d are nearly identical regardless of the RF chirp direction. The overall amplitudes of the time domain spectra are fit using the HITRAN relative line strengths and width parameters [9] in a nonlinear least-squares algorithm. The fit residuals (x4) are shown below the spectra. Evidence of some line shape error and phase drift are present in the magnified residuals, resulting in spikes and baseline drift, respectively. Additional spectra and fits are given in the supplemental S5 document for regions near 15.7 cm$^{-1}$ and 19.3 cm$^{-1}$.

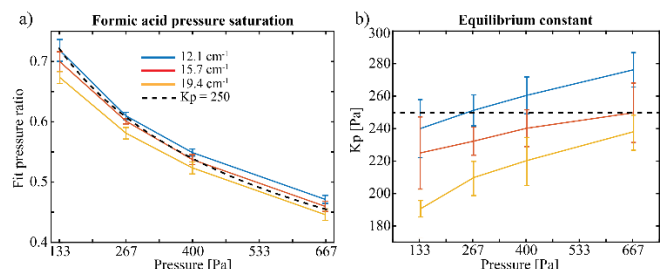

**Fig. 4**. The results of fitting the pressure of the time domain spectra acquired at 133 Pa, 267 Pa, 400 Pa and 667 Pa using fixed HITRAN relative amplitudes and width parameters. (a) The pressure dependence of the best fit pressure ratio relative to HITRAN predictions in three spectral regions. The pressure ratio for $K_P$=250 is shown as a dashed black line. (b) Fits of equilibrium constant, $K_P$, for the same data sets, showing a slight model dependence on pressure in the different regions.

Figure 4a shows the ratios of the fitted pressure to the actual pressure for three regions and at four different pressures. As the actual pressure is increased, the pressure from the fits decrease to less than half of the actual value. From previous mid-IR work [10,11], formic acid vapor at room temperature is known to exist in an equilibrium between the monomer and several possible dimeric forms. The formation of dimers results in a reduction of the monomeric partial pressure, $P_{FAM}$, and consequently, the measured line intensities as the pressure is increased. The equilibrium constant in the mid-IR region has been estimated from the pressure dependent band intensities using, $K_P = P_{FAM}^2/P_{FAD}$ where $P_{FAD}$ is the dimer partial pressure and reported to be 361(45) Pa [11]. The actual monomer pressure is defined as,

$$P_{FAM} = \frac{-K_p + \sqrt{K_p^2 + 4 K_p P_{TOT}}}{2} \quad (6)$$

where $P_{TOT}$ is the actual pressure. The sample used in our work contained 5 % water by weight and therefore, the $K_P$ is expected to change relative to the mid-IR value. As shown with a dashed black line in Fig. 4a, the predicted pressure decrease using an estimated value of $K_P$ = 250 Pa are in reasonable agreement with the observed trends.

For a more quantitative comparison, $K_P$ was used as a floated parameter in the nonlinear least squares fitting process. The fits are performed separately using the magnitude and phase spectra of each sideband spectrum and for both Rx chirped pulses. Figure 4b shows the results of the $K_P$ fits where the uncertainties were estimated from the range of values obtained for these different fits (type B, $k$=1 or 1$\sigma$). With increasing pressure, the $K_P$ values are seen to increase slightly and the overall $K_P$ values are somewhat higher for the lower frequency ranges. Further studies are needed to explain these anomalous trends which may have origin in a rotational state dependence of the rates to transient complex formation [6] affecting the equilibrium constant and/or higher order terms in the equilibrium constant associated with trimers and water complexes.

In this letter, we present a method for obtaining both the temporally magnified and steady state spectra from the down converted RF comb teeth in the frequency and time domains, respectively. The DCP-DF-EOFC system has been used to investigate the complex transmission and dispersion profiles of formic acid over three spectral regions from 362 GHz to 581 GHz. The frequency domain spectra contain distinct rapid passage effects that drastically change the appearance of the spectra depending on the defined properties of the LO chirp relative to the probe chirp. The time domain spectra of formic acid are well fit to HITRAN predictions after accounting for the gas phase equilibrium between the monomer and dimers forms at room temperature. Smaller systematic differences in the fits with increasing pressure point to a quantum state dependence of the transient complexes [6] not captured in the simple equilibrium constant model and are currently under investigation.

**Supplementary Material.** See supplemental document for supporting content.

**Acknowledgments.** J.R.S. acknowledges support from the NIST NRC fellowship program. We acknowledge helpful discussions with K. A. Briggman, J. C. Hwang and A. P. Rotunno.

**Disclosures.** The authors have no conflicts of interest to disclose. Official contribution of the National Institute of Standards and Technology; not subject to copyright in the United States.

**References**
1. S. Schiller, Opt. Lett., **27(9)**, 766 (2002).
2. I. Coddington, N. Newbury & W. Swann, Optica, **3(4)**, 414 (2016).
3. D. A. Long, A. J. Fleisher, K. O. Douglass, S. E. Maxwell, K. Bielska, J. T. Hodges & D. F. Plusquellic, Opt. Lett., **39(9)**, 2688 (2014).
4. P. Martin-Mateos, M. Ruiz-Llata, J. Posada-Roman & P. Acedo, IEEE Photonics Technology Lett., **27(12)**, 1309 (2015).
5. J. R. Stroud, J. B. Simon, G. A. Wagner & D. F. Plusquellic, Opt. Express, **29(21)**, 33155 (2021).
6. J. R. Stroud & D. F. Plusquellic, JCP, **156**, 044302 (2022).
7. E. R. Brown, Appl. Phys. Lett., **75(6)**, 769 (1999).
8. F. Bloch, Phys. Rev. **70**, 460 (1946).
9. I. E. Gordon, L. S. Rothman, C. Hill, R. V. Kochanov, Y. Tan, P. F. Bernath, ... & E. J. Zak, JQSRT, **203**, 3 (2017).
10. A. K. Roy & A. J. Thakkar, Chem. Phys., **312**, 119-126 (2005).
11. J. Vander Auwera, K. Didriche, A. Perrin & F. Keller, JCP, **126(12)**, 124311 (2007).

# A dual chirped-pulse electro-optical frequency comb method for simultaneous molecular spectroscopy and dynamics studies: Formic acid in the THz region: supplemental document

### S1. Differential chirp down conversion

The down conversion of the spectral information on the optical probe chirp into the RF domain is based on the mixing with a local oscillator (LO) chirp having the same repetition rate, but different chirp ranges. The electro-optic modulator (EOM) generates both positive and negative (±) sidebands and higher orders (harmonics) at the microwave driving field around the laser carrier frequency. While the (±) EOM sidebands are separated by the difference in acousto-optic modulator (AOM) driving frequencies, the EOM orders are more difficult to separate. The modulation of order $k$ on the probe chirp can be described as a sine wave with a quadratic phase term,

$$E^k_{probe}(t) = \widetilde{E^k}_{probe}(t) \sin\left(2\pi f_{opt}t + 2\pi k\left(f_0 t + \frac{\Delta f}{2\tau_{CP}}t^2\right)\right) \tag{S1}$$

where $f_{opt}$ is the optical carrier frequency, $f_0$ is the starting frequency of the chirp, $\Delta f$ is the chirp range of the probe, $\tau_{CP}$ is the chirp duration, and $\tilde{E}_{probe}$ is the electric field strength of the probe chirp. The LO chirp can be defined similarly as,

$$E^k_{LO}(t,i) = \widetilde{E^k}_{LO}(t) \sin\left(2\pi f_{opt}t + 2\pi k\left(f_{LO}t + \frac{\Delta f_{LO}}{2\tau_{CP}}t^2\right) - \frac{2\pi k}{N_{chirps}}\left(\frac{t}{\tau_{CP}} + i\right)\right) \tag{S2}$$

where $f_{LO}$ is the starting frequency of the LO, $\Delta f_{LO}$ is the chirp range of LO, and $\tilde{E}_{LO}$ is the electric field strength of the LO chirp. The probe chirp is repeated $N_{chirps}$ times, while the different LO chirped pulses, $i$, are phase shifted by $2\pi k$ over the $N_{chirps}$ waveforms.

In order to down convert the spectral information on the probe chirp, the combined probe and LO chirped pulses are mixed at a photomixer. The square law detector output gives us both the square of the two field amplitudes and their cross product.

$$\left(E^k_{probe}(t) + E^k_{LO}(t)\right)^2 = E^k_{probe}(t)^2 + E^k_{LO}(t)^2 + 2E^k_{probe}(t)E^k_{LO}(t) \tag{S3}$$

Using the product angle formula, we can express the product of the probe and LO chirped fields (Eqs. S1 and S2) as the difference and sum terms,

$$E^k_{probe}(t)E^k_{LO}(t) = \frac{\widetilde{E^k}_{probe}(t)\widetilde{E^k}_{LO}(t)}{2}\left(\begin{array}{l}\cos\left(2\pi k\left((f_0 - f_{LO})t + \frac{(\Delta f - \Delta f_{LO})}{2\tau_{CP}}t^2\right) + \frac{2\pi k}{N_{chirps}}\left(\frac{t}{\tau_{CP}} + i\right)\right) \\ -\cos\left(4\pi f_{opt}t + 2\pi k\left((f_0 + f_{LO})t + \frac{(\Delta f + \Delta f_{LO})}{2\tau_{CP}}t^2\right) - \frac{2\pi k}{N_{chirps}}\left(\frac{t}{\tau_{CP}} + i\right)\right)\end{array}\right) \tag{S4}$$

The high frequency components (second term in Eq. S4) are filtered at the detector, leaving the remaining difference frequency term that defines the RF field, $RF^k(t)$,

$$RF^k(t) = I^k_{RF}(t) \cos\left(\pm 2\pi k\left(\delta f_0 t + \frac{\delta f_{BW}}{2\tau_{CP}}t^2\right) + \frac{2\pi k}{N_{chirps}}\left(\frac{t}{\tau_{CP}} + i\right)\right) \tag{S5}$$

where $\delta f_0$ is the difference in starting frequencies, $\delta f_{BW}$ is the difference in chirp ranges, and $I_{RF}$ is the product of the probe and LO electric fields over 2. Eq. S5 gives a chirped pulse in the RF domain that downconverts the spectral information of the probe electric field. The RF chirp starts at $\delta f_0 = f_0 - f_{LO}$ and sweeps over the bandwidth $\delta f_{BW} = \Delta f - \Delta f_{LO}$.

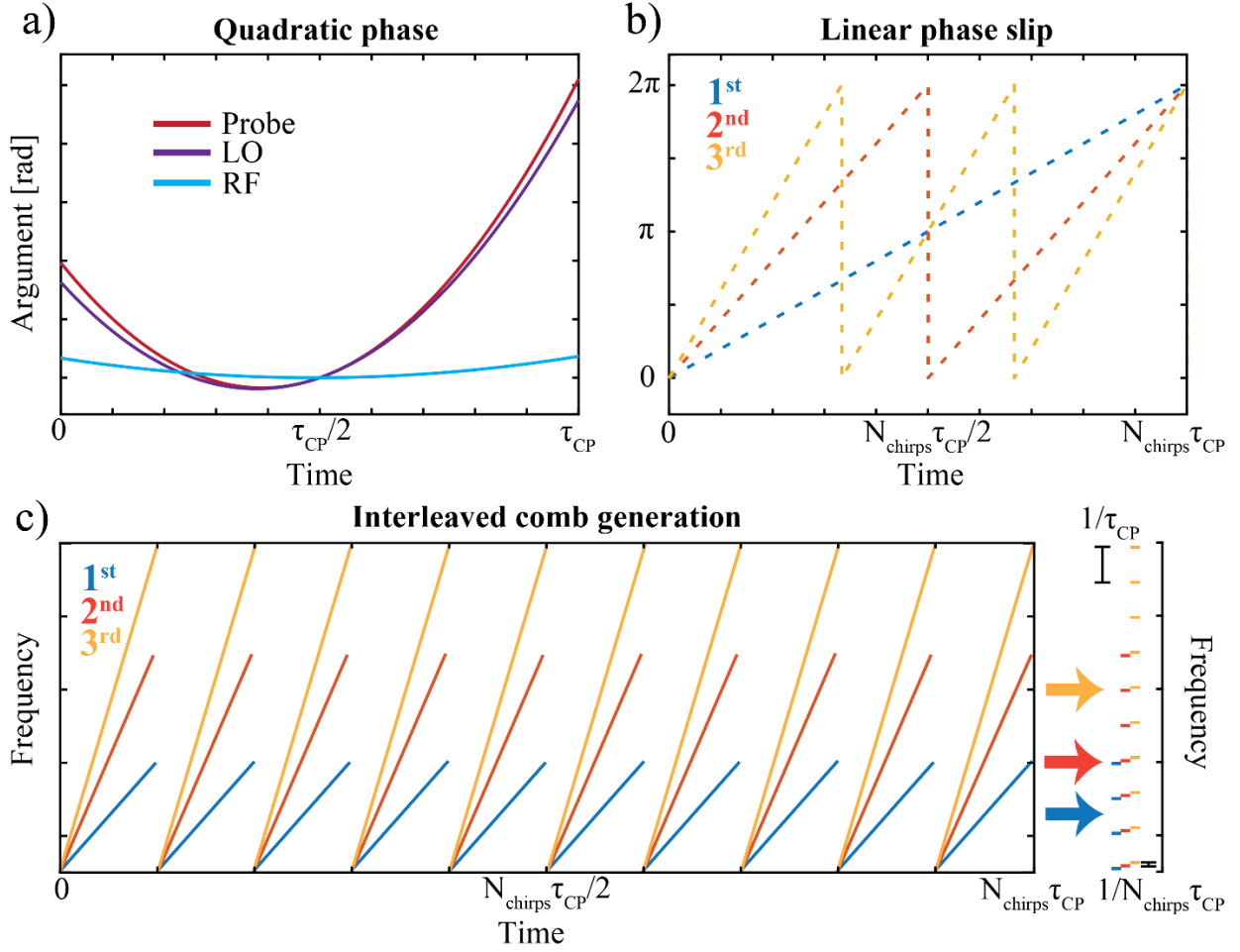

**Fig. S1**. (a) The trigometric argument of the probe chirp is shown in red, the LO chirp in purple, and the RF chirp in light blue. The two optical chirps are dominated by the quadratic phase terms that define the linear chirp ranges (the instantaneous angular frequency is the derivative of the phase), and the RF chirp range also has the quadratic shape defined by their difference. (b) The linear phase shift term that separates high orders (harmonics) of the EOMs by generating a beat note over a fixed set of chirps[1]. (c) The RF comb is generated in the frequency domain by sampling over a repeating series of chirps. The RF scan range scales with the order while the comb teeth centers over consecutive orders are sequentially separated by the linear phase shift in (b).

      The chirp is repeated to form a comb in the RF spectrum whose bandwidth is defined by the quadratic term, $\delta f_{BW}$, in Eq. S5. The sine arguments of the probe and LO chirps (i.e., Eqs. S1 and S2) are plotted in Fig S1a, in red and purple, respectively. The shapes of the quadratic forms are defined by the chirp ranges while the horizontal offsets of the minima are defined by the starting frequencies. The difference between the two quadratic functions is another quadratic (light blue) that generates the down-converted linear chirp in the RF domain. The linear frequency shift that is scaled by $N_{chirps}$ is illustrated in Fig S1b as the wrapped phase over the series of waveforms for the first three orders. Because the magnitude of the linear phase shift (or corresponding frequency shift) scales with the order of EOM modulation, the different orders are separated in the RF spectrum. The comb is formed by Fourier transforming a repeated series of chirps shown in Fig S1c, where each order spans the $k$ scaled frequency ranges for the same chirp duration. The comb has teeth that are separated by the inverse of the chirp duration, $\tau_{CP}$, while the different orders are separated by the $1/(N_{chirps}\tau_{CP})$ in the RF comb.

      For the first order, $k=1$, and ignoring the phase shift scaled by $N_{chirps}$, the RF waveform generated by mixing two chirped pulses simplifies to,

$$RF(t) = I(t)\cos\left(\pm 2\pi\left(\delta f_0 t + \frac{\delta f_{BW} t^2}{2\tau_{CP}}\right)\right) \tag{S6}$$

This is Eq. 2 in the main document.

**S2.**     **Magnified temporal dynamics**

When the probe field interacts with the sample (assuming a linear response), the output can be described as a linear combination of the transient response of the sample to each frequency in the chirp. We can first write the general single frequency transient response as,

$$E_{trans}(t) = \tilde{E}_{trans}(t)\sin(2\pi f_{trans} t) \tag{S7}$$

where $\tilde{E}_{trans}$ is the dampened envelope of the transient oscillation at frequency $f_{trans}$. When this is mixed with the LO at the detector, the resulting RF waveform contains a quadratic term defined by the LO chirp. Following Eq. S3 and Eq. S4, we retain only the difference term of the product between the transient field and the LO chirp and write,

$$RF_{trans}(t) = LPF\{E_{trans}(t)E_{LO}^k(t)\} = I_{trans}(t)\cos\left(2\pi\left(\delta f_{trans}t - \frac{k\Delta f_{LO}}{2\tau_{CP}}t^2\right) + \frac{2\pi k}{N_{chirps}}\left(\frac{t}{\tau_{CP}} + i\right)\right) \tag{S8}$$

where $\delta f_{trans}$ is the beat note between the LO and transient frequency, and $I_{trans}(t)$ is the intensity of the fields detected after low pass filtering (LPF) for the difference term. Instead of having a quadratic term that is the difference between two chirps, the spectral mapping of the transient signal is defined by the LO chirped pulse alone. When the series of repeated chirps is Fourier transformed to form a comb in the *frequency domain*, the transient responses are magnified by the quadratic phase of the LO chirp. However, when the inverse Fourier transform is applied, a linear scan response in time is recovered to reveal the normal *time domain* spectral response of the sample. This results in four different spectra recovered from each data set which include the magnitude and phase for each of the time and frequency domain spectra.

For the first order, $k=1$, and ignoring the phase shift scaled by $N_{chirps}$, the RF waveform generated by mixing a transient response and the LO chirp simplifies to,

$$RF_{trans}(t) = I_{trans}(t)\cos\left(2\pi\left(\delta f_{trans}t - \frac{\Delta f_{LO}t^2}{2\tau_{CP}}\right)\right) \tag{S9}$$

This is Eq. 3 in the main document.

## S3. Fourier transform of a linear chirp

The linear frequency chirp is defined by a quadratic phase term in a complex exponential. To give insight into the Fourier transform of a chirped pulse, it is convenient to look at the exponential form of the chirped pulse function given in Eqs. S1 and S2,

$$E_{chrip}(t) = a \, exp\left(i\left(bt + \frac{\Delta\omega}{2\tau_{CP}}t^2 + c\right)\right) \tag{S10}$$

where $a$, $b$, and $c$ are constants defining the amplitude, linear time dependence, and phase offset, respectively. In radial units ($\Delta\omega = 2\pi\Delta f$), the terms that define the quadratic dependence are the chirp range, $\Delta\omega$, and the chirp duration, $\tau_{CP}$. We can separate the linear (plus constant) and quadratic terms into,

$$E_{chrip}(t) = a \, exp(ic) \, exp(ibt) \, exp\left(i\frac{\Delta\omega}{2\tau_{CP}}t^2\right) = s(t)h(t) \tag{S11}$$

so we have both linear, $s(t)$, and quadratic, $h(t)$, functions of time. The Fourier transform of their product can be written as the Fourier transform of $h(t)$ with a frequency shift defined by $b$ in $s(t)$,

$$FT\{E_{chrip}(t)\} = FT\{s(t)h(t)\} = a \, H(\omega - b) \, exp(ic) \tag{S12}$$

The Fourier transform of the quadratic term $h(t)$ can be defined as,

$$FT\{h(t)\} = H(\omega) = \sqrt{\frac{\tau_{CP}}{\Delta\omega}} \, exp\left(i\left(-\frac{\tau_{CP}}{2\Delta\omega}\omega^2 + \frac{\pi}{4}\right)\right) \tag{S13}$$

These combine to give the full transform of the chirped pulse function,

$$FT\{E_{chrip}(t)\} = a\sqrt{\frac{\tau_{CP}}{\Delta\omega}} \, exp\left(i\left(\frac{\tau_{CP}b}{\Delta\omega}\omega - \frac{\tau_{CP}}{2\Delta\omega}\omega^2 - \frac{\tau_{CP}b^2}{2\Delta\omega} + \frac{\pi}{4} + c\right)\right) \tag{S14}$$

or more generally,

$$E_{chrip}(\omega) = A \, exp\left(i\left(B\omega - \frac{\tau_{CP}}{2\Delta\omega}\omega^2 + C\right)\right) \tag{S15}$$

where, $A$, $B$, and $C$ are constants defined in terms of the initial constants, chirp range, $\Delta\omega$, and chirp duration, $\tau_{CP}$. This simplified example results in an equation very similar in form to Eq. S10, where there is a linear term defined by the constant $B$, and a quadratic term defined by the chirp duration and chirp range. The Fourier transform of a chirp pulse is treated here in its simplest form, since any increase in the function argument complexity would require a numerical solution to solve for the transform.

This transform property of the linear chirps permits the acquisition of unique spectra in each domain where these different chirped pulse forms are represented. The quadratic phase term appearing in both Eq. S10 and Eq. S15 serve to linearly transform information from one domain to the other. Terms that are multiplied in the time domain will be convolved in the frequency domain and result in a unique spectral response depending on the time/frequency dependence of the sample response.

The directly sampled comb spectra (magnitude and phase) for a single sideband is shown in Fig. S2, with the signal spectra in blue and the background spectra in red. Figure S2a shows the sample response in the frequency domain that contains asymmetric oscillations near resonances with regions that exceed the background level. In contrast, the time domain spectra in Fig. S2b show distinct and symmetric features in transmission. The corresponding phase specta in Fig. S2c and S2d are dominated by the quadratic phase terms that enable these transforms. The quadratic phases are shifted by the linear terms, and as predicted by Eqs. S10 and S15, the sign of the overall phase is reversed in the two domains.

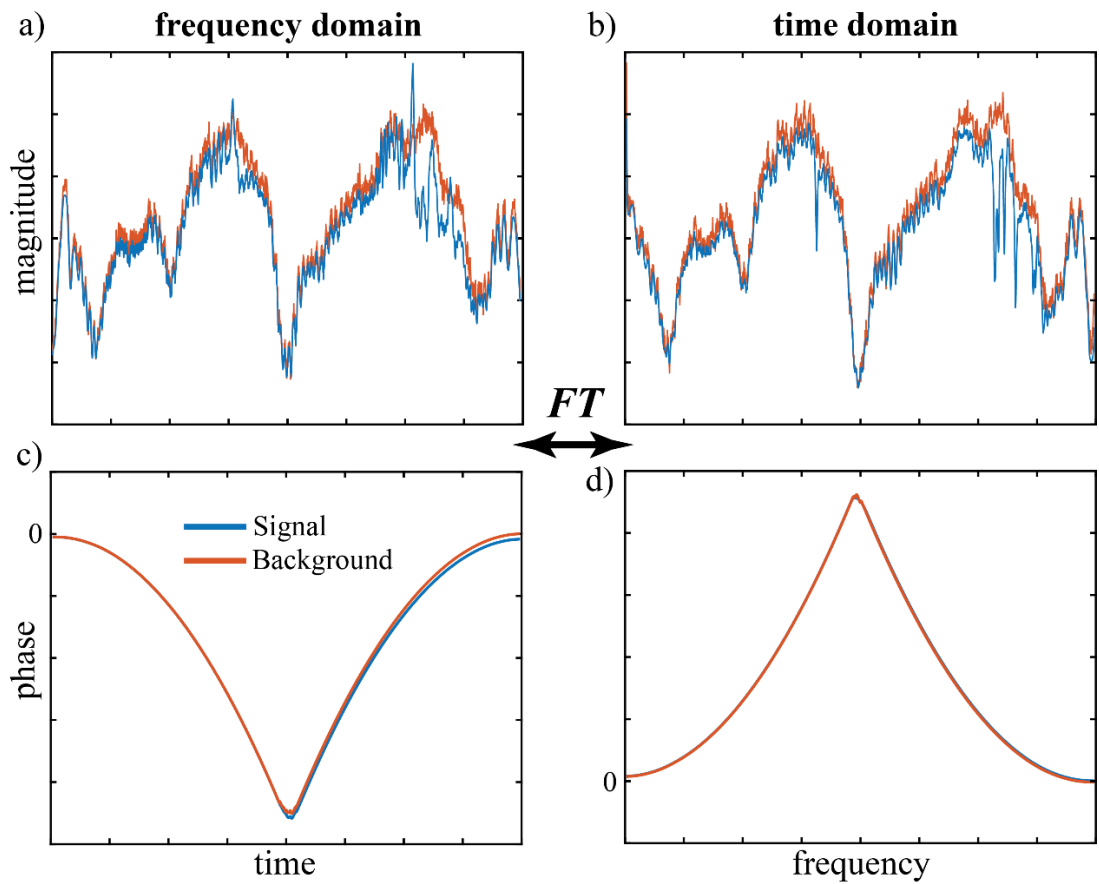

**Fig. S2**. a) The frequency domain magnitude spectra where the sampled comb lines of the signal are shown in blue, and those of the background spectra in red. b) The time domain spectra obtained from the inverse Fourier transform of the frequency domain spectra (a,c). c) The quadratic phase response of the frequency domain spectra, and d) of the time domain spectra.

## S4. Maxwell-Bloch equations: Rapid passage effects

Rapid passage effects in molecular systems have been treated in detail before using the following set of differential equations [2,3]. The Feynman-Bloch vector components, *u*, *v*, and *w*, are solved for the sweep rate, *α*. The following set of equations are solved numerically,

$$\dot{u} = -(\Delta - \dot{\phi})v - u/T_2 \qquad (S16)$$

$$\dot{v} = +(\Delta - \dot{\phi})u + \Omega w - v/T_2 \qquad (S17)$$

$$\dot{w} = \Omega v - (w - w_{eq})/T_1 \qquad (S18)$$

where $\Delta$ is the frequency detuning from resonance, $w_{eq}$ is the steady state population inversion ($w_{eq}$ = -1) and $T_1$ and $T_2$ are the population and polarization relaxation times, respectively. The rabi frequency, $\Omega$, is related to the applied electric field amplitude by $\Omega = d_{ij}E/\hbar$, where $d_{ij}$ is the transition dipole moment of the absorption line and $\hbar$ is the reduced Planck's constant. The complex envelope function $\tilde{E}(z,t) = E(z,t)\exp(-i\phi(z,t))$ describes the amplitude and phase of the electric field along the laser propagation direction, *z*.

The field at every point, z, over the pathlength, $L_{abs}$, is evaluated using Eqs. S16-S18 at discrete time steps to determine the field amplitude and phase at the detector. Averages are performed over the Boltzmann velocity components and dipole moment projections on the laser's electric field direction after accounting for degeneracies. In the retarded time reference frame where *Z* = *z* and *T* = *t* - *z/c*, the Rabi field and phase evolve in the propagation dimension with respect to the Feynman-Bloch vectors as,

$$\frac{\partial \Omega}{\partial Z} = \frac{\mu}{4}\langle v(Z,T)\rangle \qquad (S19)$$

$$\frac{\partial \phi}{\partial Z} = -\frac{\mu}{4}\frac{\langle u(Z,T)\rangle}{\Omega} \qquad (S20)$$

where $\mu = Nd_{ij}^2\omega_0/\epsilon_0\hbar c$ and *N* is the number density, $\omega_0$ is the center frequency of the absorption line, and $\epsilon_0$ is free space permittivity.

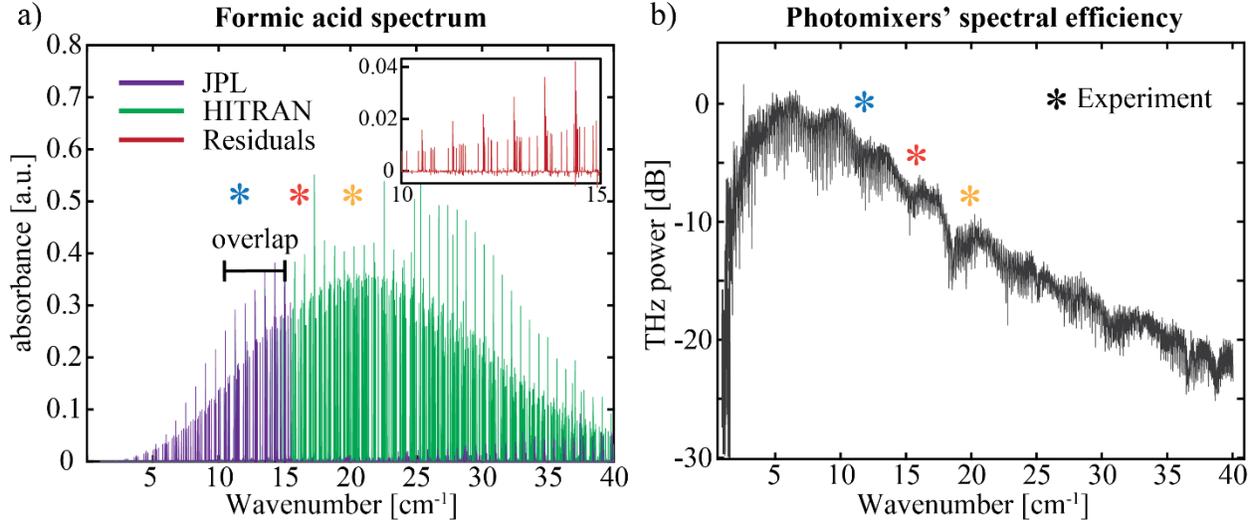

**Fig. S3**. (a) Models of the THz spectrum of formic acid from 1 cm$^{-1}$ to 40 cm$^{-1}$. The HITRAN database, in green, only reports the line parameters above 10 cm$^{-1}$, so the low frequency region is filled in with predictions from the JPL database [4], in purple. The residuals from the overlaped region are shown in the inset (red) which indicates a slightly difference in the line intensities between the two databases. (b) The power spectrum obtained from the THz photomixer pair through an evacuated 1 m long cell.

## S5. Additional data:

The dual chirped-pulse difference-frequency electro-optical frequency comb system is used to perform gas phase spectroscopy of formic acid. Figure S3a shows predictions of the absorption spectrum of the formic acid monomer from 1 cm$^{-1}$ to 40 cm$^{-1}$, or 30 GHz to 1.2 THz. The HITRAN database [5] spectrum shown in green spans down to 10 cm$^{-1}$ where it overlaps by ≈ 5 cm$^{-1}$ with the JPL database spectrum shown in purple. The difference in line intensities is illustrated by the residuals between the two spectra over the 10 cm$^{-1}$ to 15 cm$^{-1}$ region shown as a inset in red. In this manuscript, we scan over three spectral regions centered at 12.1 cm$^{-1}$, 15.7 cm$^{-1}$, and 19.3 cm$^{-1}$ (362 GHz, 472 GHz, and 581 GHz, respectively), illustrated in Fig S3 by asterisks. The measured power spectrum of the Tx and Rx photomixer pair is shown in Fig S3b. The peak power detected is near 5 cm$^{-1}$ and logarithmically decreases as the frequency is increased. Over the three spectral ranges probed, the power drops by roughly four-fold, resulting in a corresponding decrease in signal to noise (SNR). For each region, two sets of data were collected at sample pressures of 133 Pa, 267 Pa, 400 Pa, and 667 Pa, with the only difference being the chirp parameters used for the Rx comb. Each data set produces four different spectra which includes the magnitude and phase in both the frequency and time domains. The frequency domain spectra show unique line shapes that depend on the parameters of the Rx comb and dampen with increased pressure, while the time domain spectra show the normal evolution of a pressure-broadened Voigt line shape with increasing pressure.

Following the measurements at 12.08 cm$^{-1}$ as illustrated in Fig. 3, the ECDL laser was tuned to approximately 11762 cm$^{-1}$, centering the THz spectrum near 15.72 cm$^{-1}$ (472 GHz). Due to the reduced efficiency at both the Tx and Rx photomixers, the SNR is reduced by roughly two-fold for this spectral region. Figure S4a shows the magnitude and phase of the temporally magnified spectra showing the direction of rapid passage response moving towards the comb's center frequency [1,6]. In contrast, the other Rx chirp produces the rapid passage response shown in Fig. S4b where the damping oscillations move away from the center frequency. The time domain spectra in Figs. S4c and S4d correspond to the frequency domain spectra shown above them in Figs. S4a and S4b, respectively. The unscaled residuals from HITRAN fits (offset for clarity) are scaled by two and shown below in orange.

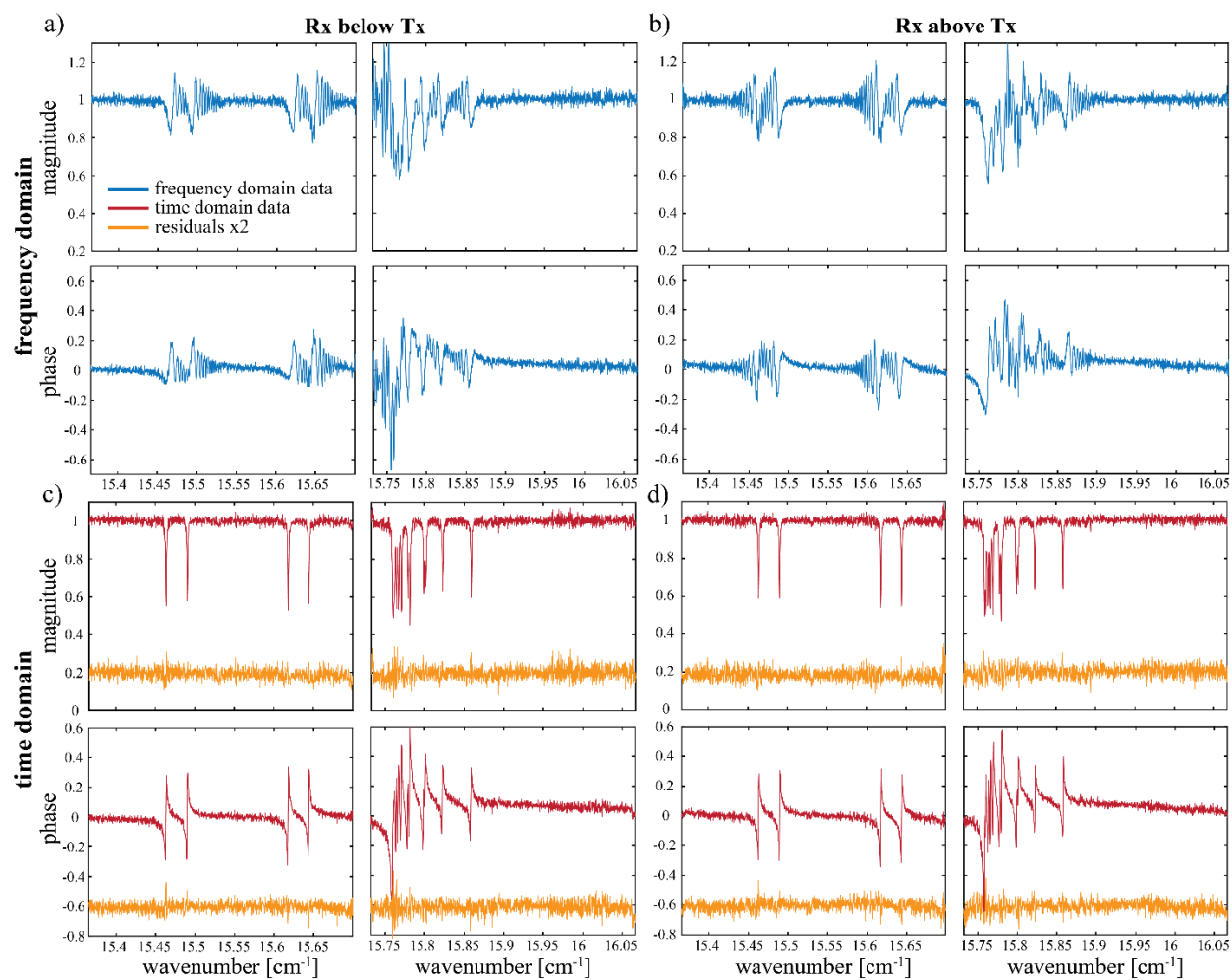

**Fig. S4**. The magnitude and phase spectra of ≈95% formic acid around 15.7 cm$^{-1}$ in a 1 m long cell at a pressure of 133 Pa. (a) The quadratic phase shift from the comb causing the ripples to move towards the center frequency. (b) The quadratic phase shift causes the ripples to move away from the laser center frequency. (c) The steady state spectra in red correspond to the temporally magnified spectra above in (a). (d) The steady state spectra in red correspond to the temporally magnified spectra above in (b), and the residuals from the HITRAN fits (offset for clarity) are shown below in orange.

The ECDL laser was then tuned to approximately 11758 cm$^{-1}$, centering the THz spectrum around 19.35 cm$^{-1}$ (581 GHz), and resulting in another reduction in SNR by a factor of two. Figure S5 shows the formic acid spectra at 267 Pa pressure. Figure S5a illustrates the magnitude and phase of the temporally magnified spectra showing the rapid passage chirp direction moving towards the comb's center frequency. The other Rx chirp produces the rapid passage chirp direction shown in Fig S5b, moving away from the center frequency. The time domain spectra in Figs. S5c and S5d correspond to the frequency domain spectra above them in Fig. S5a and S5b, respectively. Residuals from HITRAN fits are shown below in orange (offset for clarity).

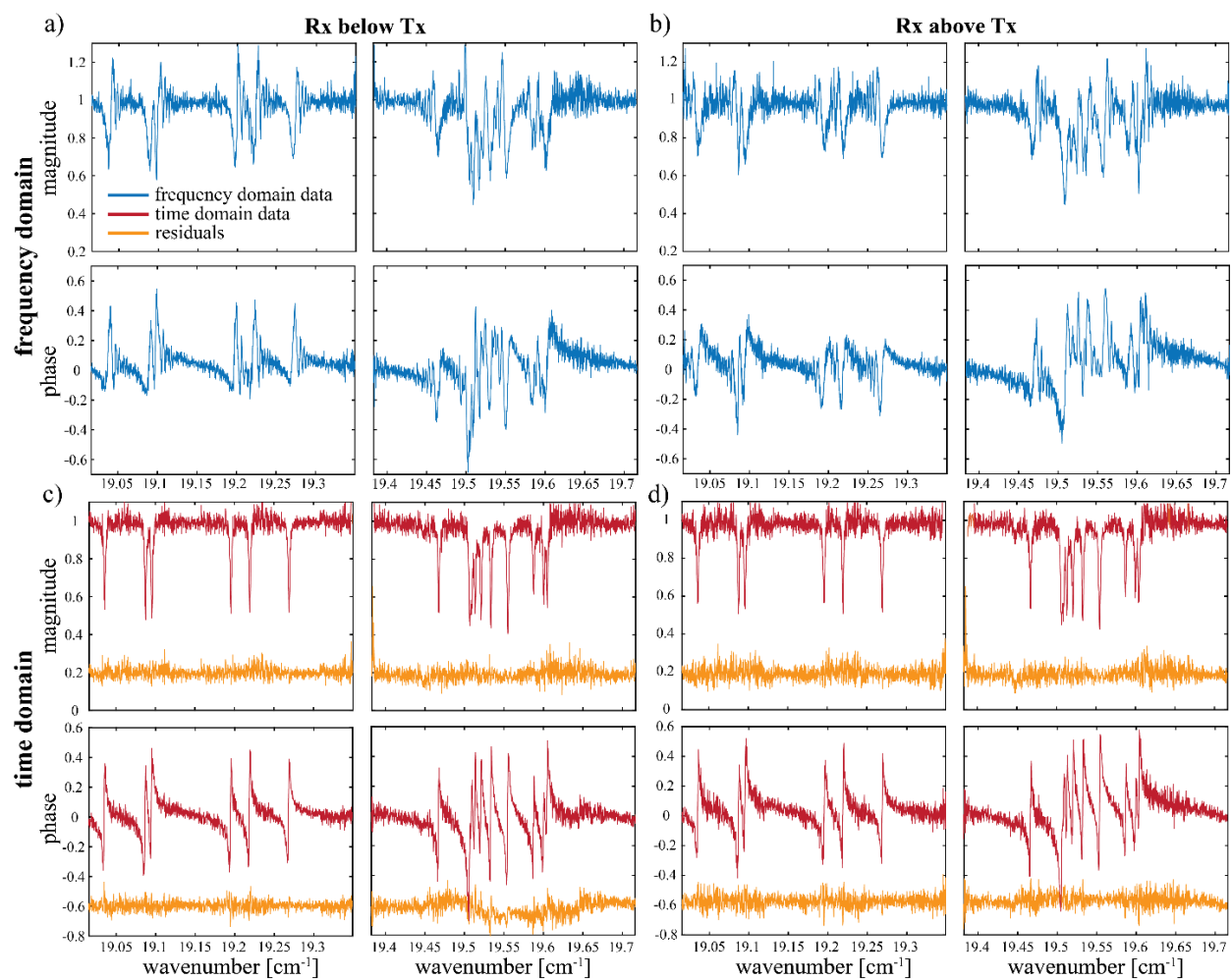

**Fig. S5**. The magnitude and phase spectra of ≈ 95% formic acid around 19.35 cm$^{-1}$ in a 1 m long cell at a pressure of 267 Pa. (a) The quadratic phase shift from the comb causing the ripples to move towards the center frequency. (b) The quadratic phase shift causes the ripples to move away from the laser center frequency. (c) The steady state spectra in red correspond to the temporally magnified spectra above in (a). (d) The steady state spectra in red correspond to the temporally magnified spectra above in (b), and the residuals to the HITRAN fits (offset for clarity) are shown below in orange.

**References:**

1 R. Stroud, J. B. Simon, G. A. Wagner & D. F. Plusquellic, Opt. Express, **29(21),** 33155 (2021).

² F. Bloch, Phys. Rev. **70**, 460 (1946).

3 N. Tasinato, K. G. Hay, N. Langford, G. Duxbury and D. Wilson, "Time dependence measurements of nitrous oxide and carbon dioxide collisional relaxation processes by a frequency down-chirped quantum cascade laser: Rapid passage signals and the time dependence of collisional processes", J. Chem. Phys. 132, 164301-(1-11) (2010).

4 Home, J. P. L. "JPL Spectral Library."

⁵. Gordon, I. E., Rothman, L. S., Hill, C., Kochanov, R. V., Tan, Y., Bernath, P. F., ... & Zak, E. J.. The HITRAN2016 molecular spectroscopic database. Journal of Quantitative Spectroscopy and Radiative Transfer, 203, 3-69 (2017).

6 J. R. Stroud & D. F. Plusquellic, JCP, 156, 044302 (2022).